\documentclass[manuscript]{aastex}

\usepackage{amsmath}

\shorttitle{PREDICTIONS FOR THE FIRST PSP ENCOUNTER}
\shortauthors{van der Holst et al.}

\begin{document}

\title{PREDICTIONS FOR THE FIRST PARKER SOLAR PROBE ENCOUNTER}

\author{B. van der Holst\altaffilmark{1}, W.B. Manchester IV\altaffilmark{1},
K.G. Klein\altaffilmark{2}, J.C. Kasper\altaffilmark{1}}

\altaffiltext{1}{Climate and Space Sciences and Engineering,
  University of Michigan, Ann Arbor, MI 48109, USA}
\altaffiltext{2}{Lunar and Planetary Laboratory,
  University of Arizona, Tucson, AZ 85719, USA}

\begin{abstract}

We examine Alfv\'en Wave Solar atmosphere Model (AWSoM) predictions of the first
Parker Solar Probe (PSP) encounter. We focus on the 12-day closest approach
centered on the 1st perihelion. AWSoM (van der Holst et al., 2014) allows us
to interpret the PSP data in the context of coronal heating via Alfv\'en wave
turbulence. The coronal heating and acceleration is addressed via
outward-propagating low-frequency Alfv\'en waves that are partially reflected by
Alfv\'en speed gradients. The nonlinear interaction of these counter-propagating
waves results in a turbulent energy cascade. To apportion the wave dissipation
to the electron and anisotropic proton temperatures, we employ the results of
the theories of linear wave damping and nonlinear stochastic heating as
described by Chandran et al. (2011). We find that during the first encounter,
PSP was in close proximity to the heliospheric current sheet (HCS) and in the
slow wind.  PSP crossed the HCS two times, namely at 2018/11/03 UT 01:02 and
2018/11/08 UT 19:09 with perihelion occuring on the south of side of the HCS.
We predict the plasma state along the PSP trajectory, which shows a dominant
proton parallel temperature causing the plasma to be firehose unstable.

\end{abstract}

\keywords{Solar Wind --- MHD --- Sun: corona --- Sun: waves --- interplanetary
medium --- methods: numerical}

\section{INTRODUCTION}

The recently launched Parker Solar Probe (PSP, \citet{fox2016})
and upcoming launch of the Solar Orbiter \citep{muller2013}
will provide a unique opportunity to track the evolution of
structures in the inner heliosphere in unprecedented detail. PSP, with its
closest perihelion of about 9.86\,R$_{\rm Sun}$ from the Sun will directly
sample the solar corona in a region never before measured in situ.
Solar Orbiter,
with its combined suite of in-situ and remote sensing instruments, will
orbit within 0.28\,AU of the Sun, and will have opportunities to remotely
observe structures
in the corona and then sample those structures in-situ during its near
co-rotational observing periods.  PSP and Solar Orbiter will revolutionize our
understanding of the solar wind with a suite of instruments that will directly
observe the state of the thermal plasma distributions \citep{kasper2016}
and electric and magnetic fields \citep{bale2016} over a
wide range of heliographic latitude and distances. These observations will
detect fluid and kinetic waves, the associated turbulent cascade of energy,
and signatures of wave dissipation and the consequent heating of the particle
populations.

The prime PSP mission includes 24 encounters with the
Sun. For each, PSP goes from 0.25\,AU to perihelion and back in about 11 days.
The perihelion distance becomes progressively closer to the Sun as the mission
progresses due to Venus gravity assists. For these orbital parameters,
PSP passes very rapidly between solar wind structures at perihelion, to sub
co-rotational speeds at aphelion.  Interpreting the resultant data require   
simulations for each encounter to put each encounter in context to understand
the observations as PSP passes through both transitent structures and fast and
slow wind streams.

Recent advances in three-dimensional (3D) time-dependent modeling
have enabled investigating the evolution of the solar wind and
solar transients as they escape the corona and are carried into the
heliosphere (see reviews by \cite{manchester2017},
\cite{kilpua2017}, and \cite{gombosi2018}).  Extended
magnetohydrodynamic (MHD) models, specifically Alfv\'en wave driven models,
are now capable of predicting turbulent wave amplitudes in the solar wind
along with the wave reflection and dissipation rates as well as partitioning
of coronal heating among particle species. In addition, full kinetic models
coupled to the MHD models include the effects of kinetic instabilities in the
solar wind, in particular on the distribution functions of the particle
species. Together,
the PSP and SO missions, along with state-of-the-art modeling capabilities
will provide the most powerful combination of tools to address fundamental
processes in the corona and heliosphere.

At the 2018 Fall AGU meeting, several 3D global MHD model predictions 
of the first PSP encounter were presented.  Among them were results from 
the CORHEL/MAS of \citet{lionello2009}, whose coronal model was 
recently extended to include low-frequency Alfv\'en wave turbulence;
the ENLIL model of \citet{odstrcil2005}, which
is an inner heliopshere MHD model prescribed by the empirical
Wang--Sheeley--Arge (WSA) model \citep{arge2000};
the solar corona and solar wind turbulence transport and
heating model by \citet{usmanov2018}; and finally, the Alfv\'en Wave
Solar atmosphere Model (AWSoM, \citet{vanderholst2014}), which is a
solar coronal and inner heliosphere model that includes low-frequency Alfv\'en
turbulence.  Here, we present the AWSoM predictions of the first PSP
encounter.

\section{AWSOM MODEL WITH ADAPT-GONG MAPS}

The simulations are peformed with the Alfv\'en Wave Solar atmosphere Model
(AWSoM, \citet{vanderholst2014}), which is a 3D solar corona and inner
heliosphere model. AWSoM solves the single fluid magnetohydrodynamic equations
extended to include proton temperature anisotropy and isotropic electron
temperature. In this model, excess heat dissipated in the corona is
transported back via electron heat conduction to the chromosphere where it is
lost via radiative cooling.

AWSoM addresses the coronal heating and wind acceleration
via low-frequency Alfv\'en wave turbulence in which partial wave
reflection is caused by Alfv\'en speed gradients. The non-linear interaction
of these counter-propagating waves results in a transverse energy cascade from
the outer scale through the self-similar intertial range to the proton
gyro-radius scale, where the Alfvenic cascade transitions into a kinetic
Alfv\'en wave (KAW) cascade.
Dissipation at the end of the cascade results in isotropic heating of the
electrons and parallel and perpendicular proton heating. To apportion the
dissipation, we include in AWSoM both stochastic ion heating by low-frequency
Alfvenic turbulence \citep{chandran2010} as well as linear Landau and transit
time damping of KAWs, with the specific partitioning developed by
\citet{chandran2011}.

Proton temperature anisotropy instabilities, specifically the CGL
(Chew, Goldberger, and Low) Firehose \citep{chew1956},
mirror \citep{tajiri1967}, and proton-cyclotron
\citep{kennel1966} instabilities are accounted for by
adding a relaxation source term to the parallel proton
pressure equation. If the plasma exceeds this stability threshold, the 
source term will relax the plasma back to the marginal stable state
with a relaxation time that is the inverse of the growth rate of the 
instability.

For the predictions of the first PSP encounter, we use synchronic magnetic
maps
that were constructed by the Air Force Data Assimilative Photospheric flux 
Transport (ADAPT) model (\cite{henney2012}, \cite{hickmann2015})
based on the Global Oscillation Network Group (GONG) maps. The ADAPT model uses
a flux transport model to predict the magnetic field in areas where the data
are not available. ADAPT uses an ensemble of twelve model realizations based on
different model parameters.

\begin{figure}[!h]
{\resizebox{0.48\textwidth}{!}{\includegraphics[clip=]{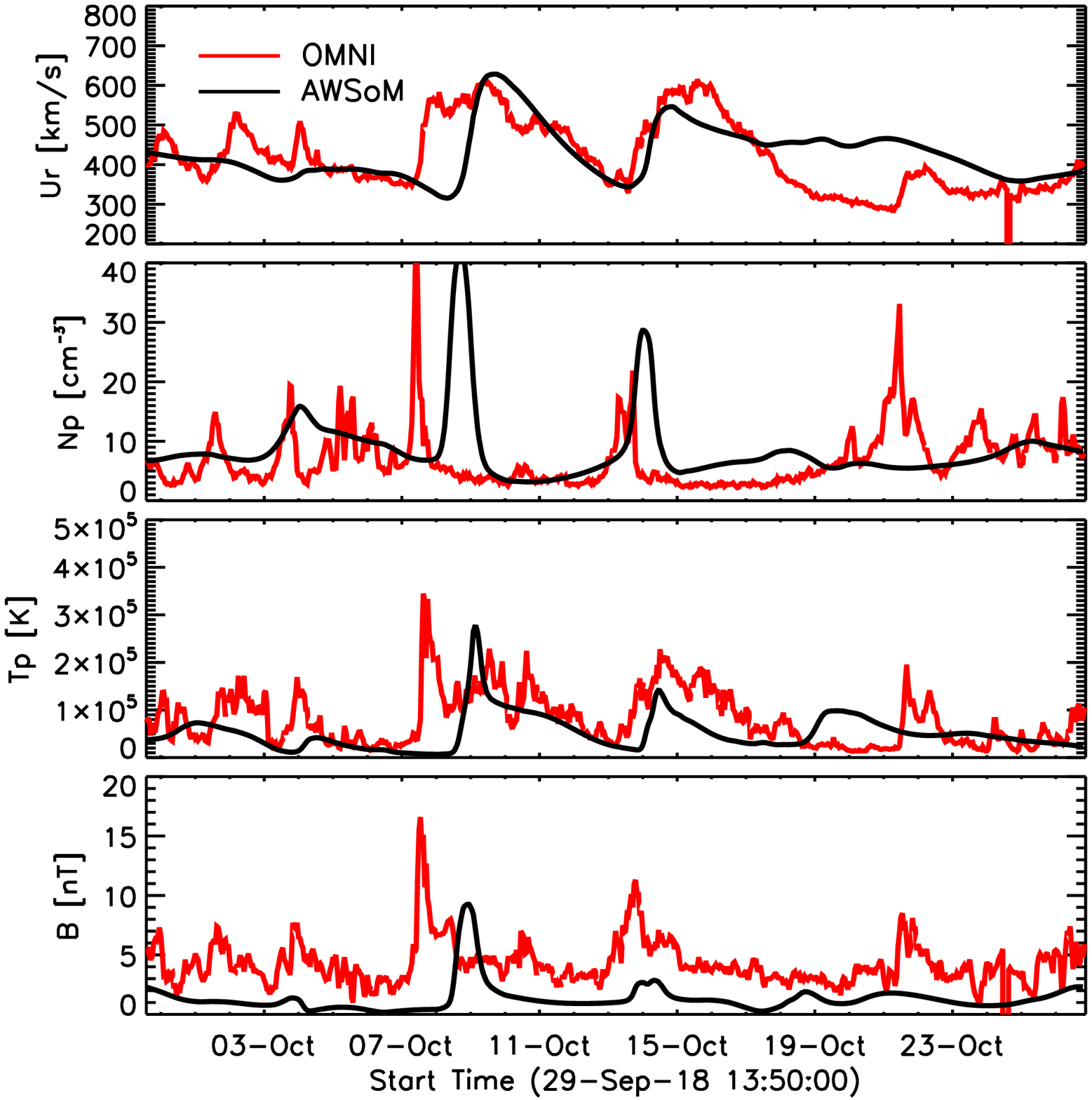}}}
\caption{Comparison of the radial solar wind speed, proton number density,
proton temperature, and magnetic field strength of the model output along
the Earth orbit with the OMNI data for CR2209.}
\label{fig:beforePSP}
\end{figure}

To determine which GONG-ADAPT model realization works best for the AWSoM
simulation of the PSP encounter, we first performed a 1AU validation
study for all twelve GONG-ADAPT maps for Carrington rotation (CR) 2209,
which was the last rotation before the PSP encounter.
Figure \ref{fig:beforePSP} displays the solar wind speed, proton number
density, proton average total temperature, and magnetic field strength along the
Earth orbit for the model (black) and OMNI data (red), which contains the
solar wind plasma and interplanetary magnetic field (IMF) conditions at 1AU. 
This simulation was performed with
a GONG-ADAPT map with central meridian time 2018/10/13 UT 06:00. In this
figure, we only show the first realization, since for this map the model
output was closest to the observed data for the last few days of CR2209.

\section{RESULTS}

The perihelion of the first PSP encounter was at 2018/11/6 UT 3:27. To
simulate the solar corona and inner heliosphere conditions for the first
encounter we used the central meridian time for the GONG-ADAPT map that is
closest to this time, hence we used 2018/11/6 UT 4:00. Since the first
realization of the ADAPT maps provided the best 1AU results for CR2209, we
used the first ADAPT realization for the PSP encounter as well.
With this map we converged the AWSoM model to a steady state for the solar
corona and inner heliopshere in the Heliographic rotating (HGR) coordinate
system.

\begin{figure}[!h]
{\resizebox{0.48\textwidth}{!}{\includegraphics[clip=]{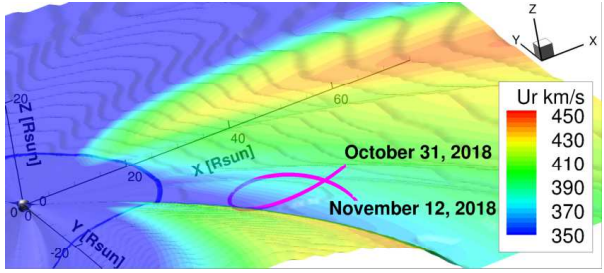}}}
\caption{The heliospheric current sheet colored with radial solar wind speed.
The magenta line is the PSP trajectory in the co-rotating frame. The $X$, $Y$,
and $Z$ coordinates in the HGR system are normalized to the solar radius.}
\label{fig:CS}
\end{figure}

In Figure \ref{fig:CS}, we show the heliospheric current sheet (HCS) for this
steady state. This isosurface, for which the radial magnetic field is zero,
is colored with the radial solar wind speed. The gray-scale sphere on the left
represents the Sun. The magenta line is the PSP trajectory of the first
encounter from October 31, 2018 to November 12, 2018 shown in the co-rotating
frame. We note that in this
time frame PSP crossed according to the AWSoM simulation two times
the HCS, namely 2018/11/03 UT 01:02 and 2018/11/08 UT 19:09.
From this figure, we see that PSP was to the south of the HCS around the
perihelion and to the north of the HCS near the beginning and end of the
encounter. During the entire encounter PSP was in close proximity to the HCS.

\begin{figure}[!h]
\begin{center}
{\resizebox{0.48\textwidth}{!}{\includegraphics[clip=]{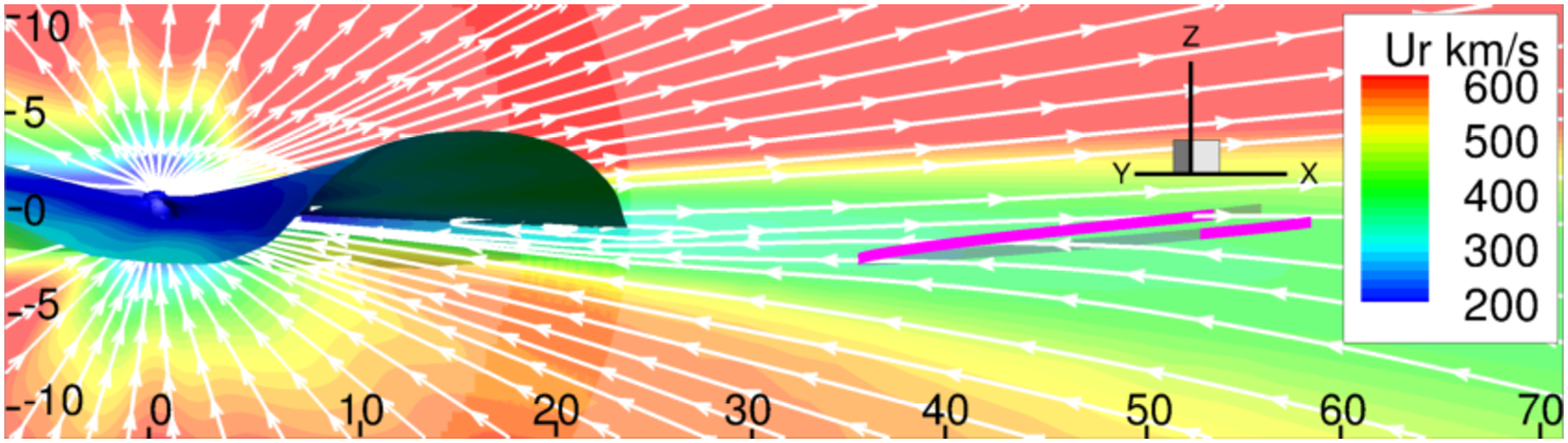}}}
{\resizebox{0.48\textwidth}{!}{\includegraphics[clip=]{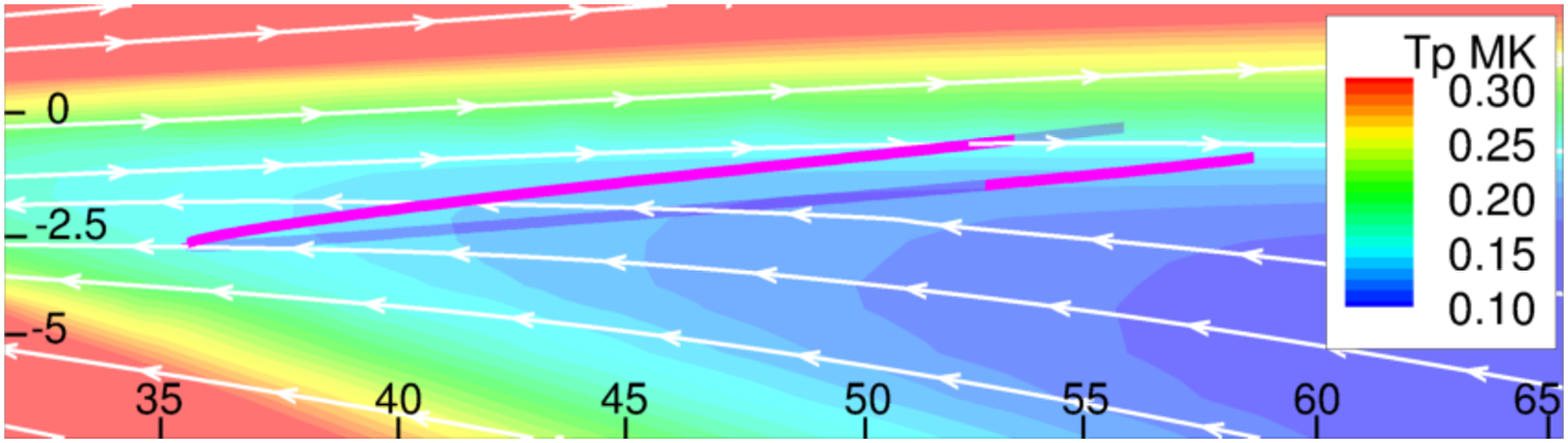}}} \\
{\resizebox{0.48\textwidth}{!}{\includegraphics[clip=]{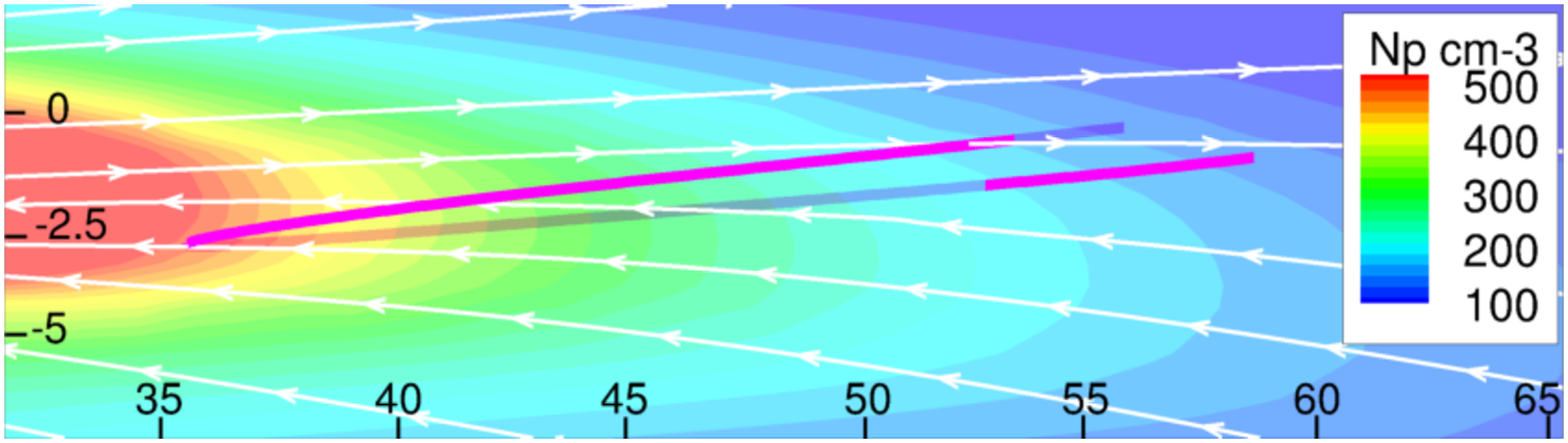}}}
{\resizebox{0.48\textwidth}{!}{\includegraphics[clip=]{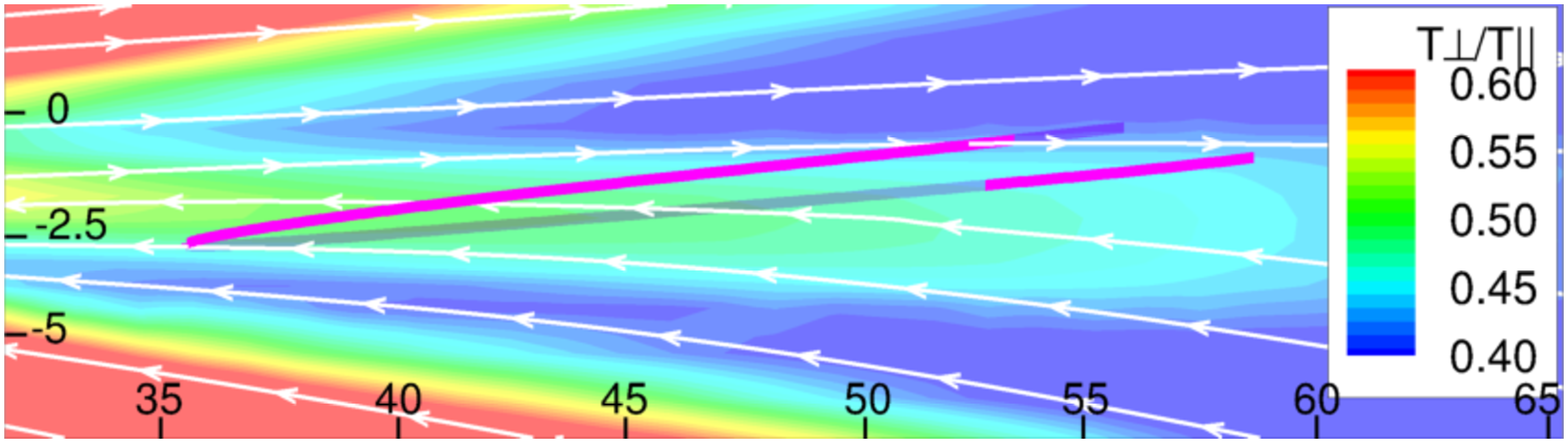}}} \\
{\resizebox{0.48\textwidth}{!}{\includegraphics[clip=]{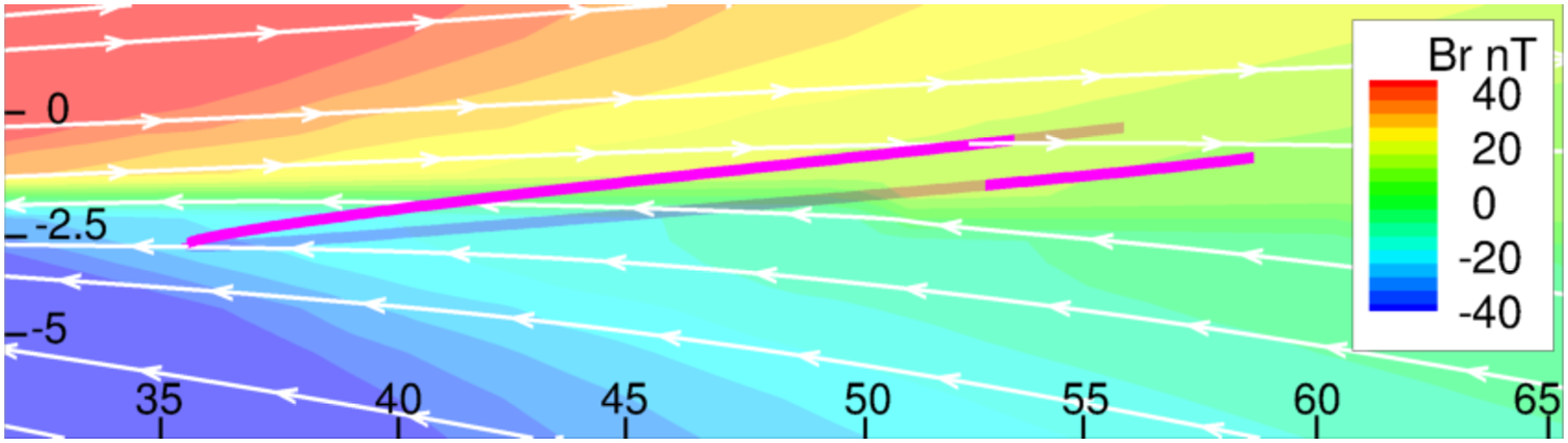}}}
{\resizebox{0.48\textwidth}{!}{\includegraphics[clip=]{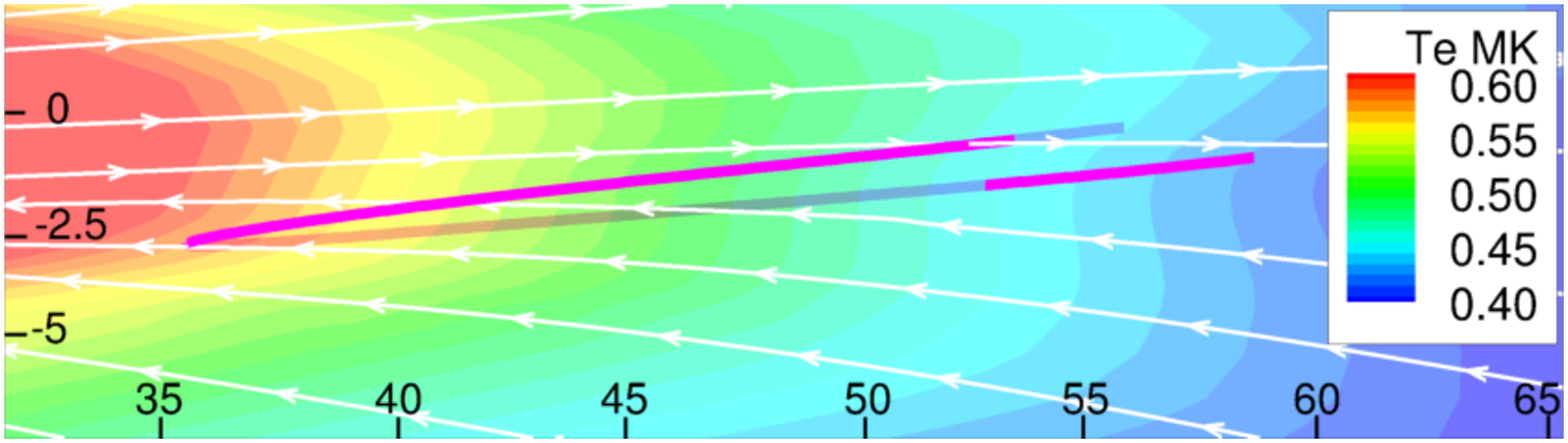}}}
\end{center}
\caption{Plasma parameters in a meridional slice through the PSP perihelion 
in the co-rotating frame.
Left panels (from top to bottom): radial velocity, proton number density, and
radial magnetic field component in color contour. Streamlines represent field
lines by ignoring the out-of-plane component. Magenta line is the PSP
trajectory. The top panel shows in addition the heliospheric current sheet for
$r < 24 R_\odot$ colored with the radial velocity. Right panels (from top to
bottom): averaged proton temperature, proton temperature anisotropy, and
electron temperature. In all panel, the vertical axis is the $Z$-direction and
the horizontal direction is the $R=\sqrt{X^2+Y^2}$ direction in solar radius.}
\label{fig:meridional}
\end{figure}

In Figure \ref{fig:meridional}, we show various plasma parameters in a section
of a meridional slice through the Sun center and the PSP perihelion in the
co-rotating frame. The top
left panel shows the radial velocity in color contour. Streamlines represent
field lines by ignoring the out-of-plane component. The trajectory of the
PSP encounter is shown as a magenta line. From this we conclude that the PSP
was close to the HCS and in the slow wind. The middle left panel shows that
the proton density went to almost 500\,cm$^{-3}$ and the bottom left panel
shows that the radial field strength during the encounter is below
30\,nT. The temperatures are shown in the right panels.  During the entire
encounter PSP would, according to the AWSoM model, detect a proton temperature
anisotropy $T_\perp/T_\parallel$ that is below 0.6.
This is a direct consequence of the significant
parallel proton heating in the AWSoM model near the HCS where the plasma beta
is high. Overall the average proton temperature (top right panel) is lower
near the HCS than the electron tempeture (bottom right panel).

\begin{figure}[!h]
\begin{center}
{\resizebox{0.48\textwidth}{!}{\includegraphics[clip=]{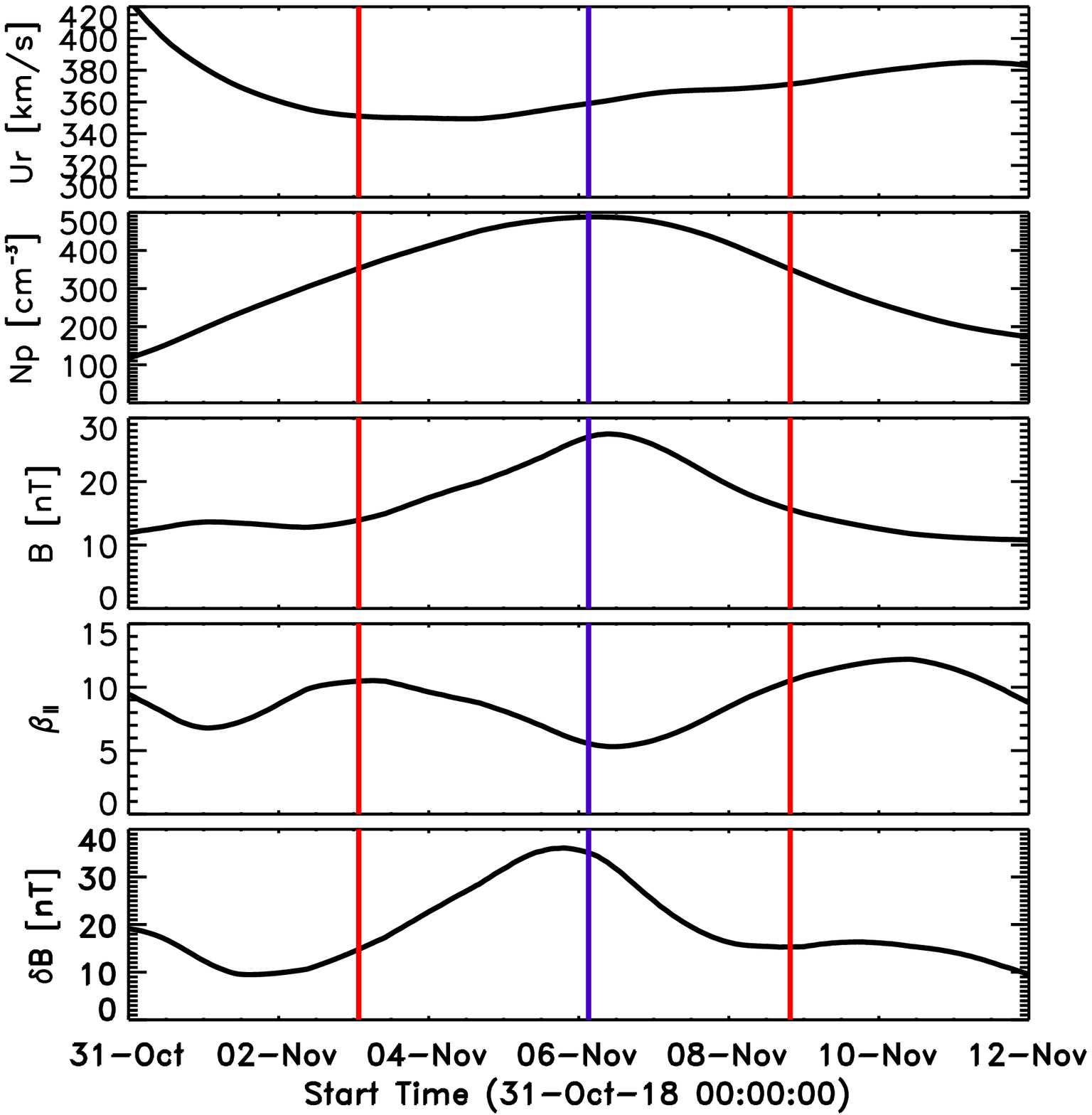}}}
{\resizebox{0.48\textwidth}{!}{\includegraphics[clip=]{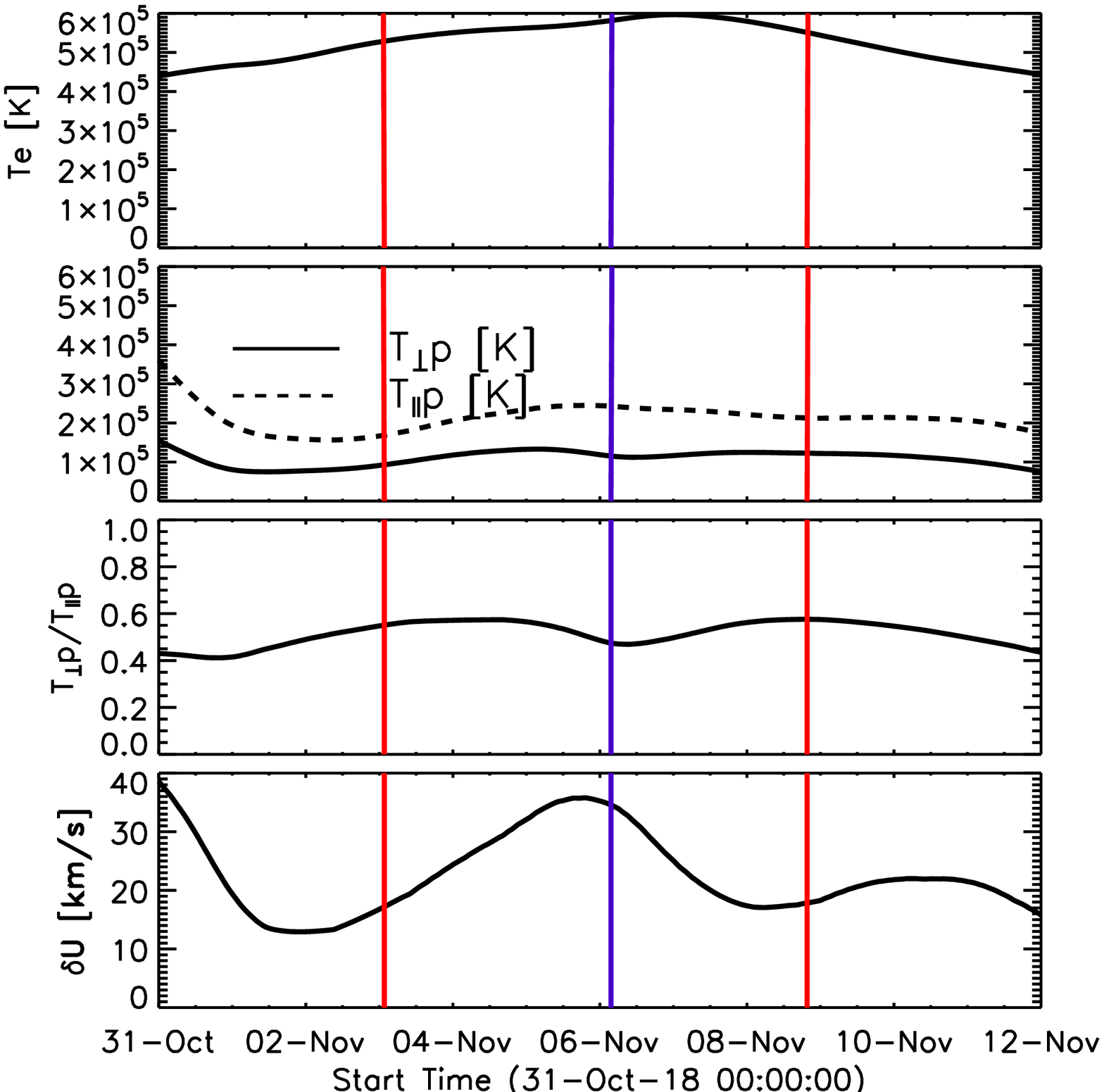}}}
\end{center}
\caption{The simulated plasma parameters during the first PSP encounter.
Left panels (from top to bottom): radial velocity, proton number density,
magnetic field strength, parallel proton plasma beta $\beta_\parallel$,
 amplitude of Alfv\'en wave magnetic field fluctuation.
Right panels (from top to bottom): electron temperature, parallel and
perpendicular proton temperatures, proton temperature anisotropy, amplitude
of Alfv\'en wave velocity perurbation. Blue vertical line indicates PSP
perihelion and red lines indicate heliospheric current sheet crossing.}
\label{fig:plasma}
\end{figure}

In Figure \ref{fig:plasma}, we show plasma parameters as a function of
time along the simulated
PSP trajectory. In this figure, the PSP perihelion is indictated by a blue
vertical line and the two HCS crossings are indicated by red vertical lines.
The top four panels on the left are for the radial velocity,
proton number density, magnetic field strength, and parallel proton
plasma beta $\beta_{\parallel p}=2p_\parallel\mu_0/B^2$, respectively.
At perihelion, AWSoM predicts $U_r = 360\,$km/s, $N_p = 490\,$cm$^{-3}$,
$B = 27\,$nT, and $\beta_{\parallel p}=5.56$ for these quantities.
The top three panels on the right
are for the electron temperature, parallel and perpendicular proton
temperature, and proton temperature anisotropy ratio (shown top to bottom). 
For the temperatures, AWSoM
predicts at perihelion $T_e = 0.58\,$MK, $T_{\perp p} = 0.12\,$MK,
and $T_{\parallel p} = 0.24\,$MK. AWSoM uses the Alfv\'en wave energy densities
parallel and antiparallel to the magnetic field lines
to calculate the impact of wave turbulence on the solar corona and inner
heliopshere plasma. To translate these energy densities into velocity and
magnetic field fluctuations, we make an additional assumption that the
reflected wave energy density propagating back to the Sun is small compared
to the outward wave energy density. These fluctuations are shown in the bottom
panels and we obtain at perihelion $\delta B = 35\,$nT and
$\delta U = 35\,$km/s.

\begin{figure}[!h]
{\resizebox{0.96\textwidth}{!}{\includegraphics[clip=]{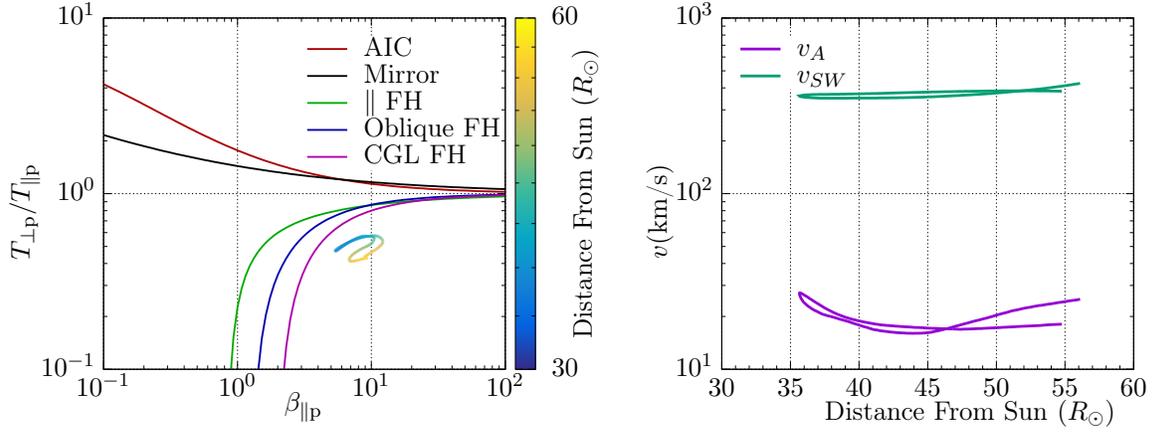}}}
\caption{Left panel: PSP trajectory in the $(\beta_{\parallel p}, T_{\perp p}/
T_{\parallel p})$--plane. The plasma along the trajectory is firehose
unstable. Right panel: radial solar wind speed and Alfv\'en speed as a
function of the radial distance of PSP.}
\label{fig:firehose}
\end{figure}

From the third panel on the right in Figure \ref{fig:plasma}, we see that
the proton temperature anisotropy $T_\perp/T_\parallel$
stays below 0.6 along the PSP
trajectory, which together with the small magnetic field strength and hence
high plasma beta result in a firehose unstable plasma. In Figure
\ref{fig:firehose}, we show the trajectory in the
$(\beta_{\parallel p}, T_{\perp p}/T_{\parallel p})$-plane. Here, we also
show the curves for which the proton-cycotron, Mirror, CGL firehose,
parallel firehose, and oblique firehose are marginally stable.
Values for these thresholds are taken from \cite{verscharen2016}.
Indeed, the
PSP trajectory is located below the firehose curves, suggesting that PSP was,
during the encounter, in firehose unstable plasma. We note that AWSoM currently
only incorporates the CGL version of the firehose instability. The parallel
and oblique firehose instabilities are planned in future updates of AWSoM and
will likely result in less anisotropy in the proton temperatures. The right
panel in Figure \ref{fig:firehose} shows the radial solar wind speed and
Alfv\'en speed ($v_A$) versus the radial distance of PSP. The model predicts
that the solar wind remains super Alfvenic over the entire trajectory of the
first encounter.

\section{SUMMARY}

In this paper, we made predictions of the first PSP encounter with the AWSoM
global solar corona and inner heliosphere model. We find that PSP was close
to the HCS, the trajectory was in the slow wind, the wind speed remains
super Alfvenic along the entire encounter trajectory,  and the plasma along the
trajectory was firehose unstable. PSP crossed the HCS two times:
2018/11/03 UT 01:02 and 2018/11/08 UT 19:09. At perihelion, PSP was to the
south of the HCS. At perihelion, AWSoM predicts for the plasma state:
$U_r = 360\,$km/s, $N_p = 490\,$cm$^{-3}$, $B = 27\,$nT, $T_e = 0.58\,$MK,
$T_{\perp p} = 0.12\,$MK, $T_{\parallel p} = 0.24\,$MK,
$\beta_{\parallel p}=5.56$, $v_A = 26.9\,$km/s, $\delta B = 35\,$nT,
and $\delta U = 35\,$km/s. In future work, we will compare the predictions
with the PSP data and make model improvements to make those comparisons
better.

\acknowledgments

This work was supported by the NSF grant 1663800 and NASA grants NNX16AL12G
and NNX17AI18G.
The simulations were performed on the NASA Advanced Supercomputing system
Pleiades.
This work utilizes data produced collaboratively between AFRL/ADAPT and
NSO/NISP.

\null\newpage

\null\newpage

\null\newpage

\newpage

\newpage

\begin{thebibliography}{}

\bibitem[Arge \& Pizzo (2000)]{arge2000}
  Arge, C.N. \& Pizzo, V.J. 2000, \jgr, 105, 10465
\bibitem[Bale et al.\ (2016)]{bale2016}
  Bale, S.D., Goetz, K., Harvey, P.R., et al. 2016, \ssr, 204, 49
\bibitem[Chandran et al.\ (2010)]{chandran2010}
  Chandran, B.D.G., Li, B., Rogers, B.N., Quataert, E., Germaschewski, K.
  2010, \apj, 720, 503
\bibitem[Chandran et al.\ (2011)]{chandran2011}
  Chandran, B.D.G., Dennis, T.J., Quataert, E., \& Bale, S.D.
  2011, \apj, 743, 197
\bibitem[Chew et al. (1956)]{chew1956}
  Chew, G.F., Goldberger, M.L., Low, F.E. 1956,
  Proceedings of the Royal Society of London. Series A, Mathematical and
  Physical Sciences, 236, 112
\bibitem[Fox et al.\ (2016)]{fox2016}
  Fox, N.J., Velli, M.C., Balse, S.D. 2016, \ssr, 204, 7
\bibitem[Gombosi et al.\ (2018)]{gombosi2018}
  Gombosi, T.I., van der Holst, B., Manchester, W.B., \& Sokolov, I.V. 2018,
 Living Reviews in Solar Physics, 15, 4
\bibitem[Kilpua et al.\ (2017)]{kilpua2017}
  Kilpua, E., Koskinen, H.E.J., \& Pulkkinen, T.I. 2017,
  Living Reviews in Solar Physics, 14, 5 
\bibitem[Henney et al.\ (2012)]{henney2012}
  Henney, C.J., Toussaint, W.A., White, S.M., \& Arge, C.N. 2012, space
  weather, 10, S02011
\bibitem[Hickmann et al.\ (2015)]{hickmann2015}
  Hickmann, K.S., Godinez, H.C., Henney, C.J., \& Arge, C.N. 2015, \solphys,
  290, 1105
\bibitem[Kasper et al.\ (20160)]{kasper2016}
  Kasper, J.C., Abiad, R., Austin, G., et al. 2016, \ssr, 204, 131
\bibitem[Kennel et al. (1966)]{kennel1966}
  Kennel, C.F. \& Petschek, H.E. 1966, \jgr, 71, 1
\bibitem[Manchester et al.\ (2017a)]{manchester2017}
  Manchester, W., Kilpua, E.K.J., Liu, Y.D., Lugaz, N., Riley, P., 
  T\"or\"ok, T., \&  Vr{\v s}nak, B. 2017, \ssr, 212, 1159
\bibitem[M\"uller et al.\ (2013)]{muller2013}
  M\"uller, D., St. Cyr, O.C. 2013, Proc.SPIE, 8862, 8862
\bibitem[Odstrcil et al.\ (2005)]{odstrcil2005}
  Odstrcil, D., Pizzo, V.J., \& Arge, C.N. 2005, \jgr, 110, A02106
\bibitem[Lionello et al.\ (2009)]{lionello2009}
  Lionello, R., Linker, J.A., \& Miki\'c, Z. 2009, \apj, 690, 902
\bibitem[Tajiri (1967)]{tajiri1967}
  Tajiri, M. 1967, Journal of the Physical Society of Japan, 22, 1482
\bibitem[Usmanov et al.\ (2018)]{usmanov2018}
  Usmanov, A.V., Matthaeus, W.H., Goldstein, M.L., \& Chhiber, R. 2018,
  \apj, 865,25
\bibitem[van der Holst et al.\ (2014)]{vanderholst2014}
  van der Holst, B., Sokolov, I.V., Meng, X., et al. 2014, \apj, 782, 81
\bibitem[Verscharen et al. (2016)]{verscharen2016}
  Verscharen, D., Chandran, B.D.G., Klein, K.G., Quataert, E. 2016,
  \apj, 831, 128
\end{thebibliography}
\end{document}